\documentclass[12pt]{article}
\usepackage{eurosym}
\usepackage{graphicx}
\usepackage[utf8]{inputenc}
\usepackage[T1]{fontenc}
\usepackage{indentfirst}
\usepackage[margin=0.6in,nomarginpar]{geometry}
\usepackage[final]{hyperref}
\usepackage{amsmath}
\usepackage{hyperref}
\usepackage{cite}
\usepackage{subcaption}
\usepackage{caption}
\usepackage{amssymb}
\usepackage{multirow}

\hypersetup{
colorlinks=true,
linkcolor=blue,
citecolor=blue,
filecolor=magenta,
urlcolor=blue
}

\begin{document}

\title{Quantum-corrected Schwarzschild AdS black hole surrounded by quintessence\ : Thermodynamics and Shadows }
\author{B. Hamil\thanks{%
hamilbilel@gmail.com, bilel.hamil@edu.umc.dz } \\
%EndAName
Laboratoire de Physique Math\'{e}matique et Subatomique, Facult\'{e}\ des
Sciences Exactes, \\
Universit\'{e}\ Constantine 1, Constantine, Algeria. \and B. C. L\"{u}tf\"{u}%
o\u{g}lu\thanks{%
bekir.lutfuoglu@uhk.cz (Corresponding author)} \\
%EndAName
Department of Physics, University of Hradec Kr\'{a}lov\'{e},\\
Rokitansk\'{e}ho 62, 500 03 Hradec Kr\'{a}lov\'{e}, Czechia. \and L.Dahbi%
\thanks{%
l.dahbi@ens-setif.dz} \\
%EndAName
Teacher Education College of Setif, Messaoud Zeghar, Algeria}
\date{\today }
\maketitle

\begin{abstract}
In this manuscript, we study the quantum-corrected Schwarzschild AdS black hole surrounded by quintessence matter in two folds: its thermal quantities and shadows. For this purpose, first, we present a detailed analysis of the influences of the quintessence matter field on Hawking temperature, mass, and specific heat functions.  Then, we examine the shadows by following Carter's approach. We find that the shadow radius value takes smaller values for greater quintessence state parameter values. On the other hand, we see that the shadow radius takes greater values for greater values of the normalization parameter. Finally, in a brief section, we discuss the energy emission rate and show that the Gaussian-type plots' peak values are also affected by the quintessence matter field parameters.
\end{abstract}

%EndAName

\section{Introduction}

Black holes are undoubtedly one of the most mysterious objects in the universe. According to Einstein's theory of general relativity, a sufficiently compact mass can deform spacetime to embody a black hole, where nothing, including light or other electromagnetic waves, can run away because of the strong gravitational effects. This belief began to change after Hawking's revolutionary work \cite{Hawkink}, in which he predicted the possibility of emitting radiation. Indeed, in the same years, Bekenstein's interpretation, which relates black hole entropy to the event horizon area, led to a complete change in the classical understanding of black holes, and hence, caused a new field of research: black hole thermodynamics. Since then, many papers have been published to comprehend and interpret the laws of black hole thermodynamics extensively \cite{Hawking75, Aman2003, Dolan2011, Carlip2014, Mann2015, Kumar2020}.

Mathematically, black holes are described by solutions of Einstein's field equation. One of the best-known was given by Schwarzschild in 1916 to identify a non-rotating black hole. Since then, Schwarzschild's black hole (SBH) \cite{Sc1, Sc2, Sc3, Sc4, Sc5, Sc6, Sc7, Sc8} and its thermodynamics \cite{Sct1, Sct2, Sct3, Sct4, Sct5, Sct6, Sct7, Sct8} attracted great interest. However, the SBH metric has an important defect: singularity problem at its two horizon radii. Thankfully, one of them is just a coordinate singularity and it can be removed with a proper coordinate transformation, hence, it has no physical meaning. Unfortunately, the other one is a physical singularity that cannot vanish simply. This intrinsic singularity prevents a perfect background structure and leads to concerns in causality \cite{1,2}. In the literature, several approaches are employed to address this issue \cite{I1, 10, I3, I2, B, C, D, 3, 5, 6, 7, 9, Sudhaker, EPL23}. In one of them, Kazakov and Solodukhi proposed to consider spherically symmetric quantum fluctuations of the metric within the Einstein-Hilbert action \cite{10}. By doing so, they eliminated the intrinsic singularity and obtained the metric of the Kazakov-Solodukhin black hole, which has a central 2D sphere with a radius of the order of the Planck length. Naturally, this change in the black holes' line element alters the geometry of the event horizon and hence the thermodynamic properties of the black hole \cite{11,12,14,15,16,18,19,20,21,22,23,24,24b,25,26, 27, Milad, Milad2, Sakalli, delgado, panella, BCLPDU, Morais}.

According to recent observations on the magnitude-redshift relation of astronomical objects, our universe is expanding at an accelerating rate \cite{Riess1, Riess2, Perlmutter1999}. The theoretical value of Einstein's cosmological constant, which is defined to explain this phenomenon, is very different from the experimental results. Therefore, physicists have been searching for new ways to explain this phenomenon. One of the most popular proposals is the concept of dark energy with a negative pressure distributed relatively homogeneously in space \cite{Peebles}. To model the dark energy various modified matter and gravity fields are being used \cite{Yoo}. One of these models is the quintessence matter model which is described by dynamical scalar fields with a characteristic state parameter that indicates the ratio between the pressure and energy density of dark energy \cite{carroll1998, Khoury2004, Picon2000, Padman2002, Caldwell2002, Gasperini2002, Copeland2006}. Keeping these facts in mind Kiselev explored the thermal quantities of the SBH surrounded by the quintessence matter field two decades ago in \cite{Kiselev}. Then, similarly, the thermal properties of other black holes that are surrounded by the quintessence matter fields have been discussed in detail in \cite{Wei2011, Thomas, XuKerr, Salwa, Ghaderi, Ghaderi16, Liu2019, Ndongmo, Zhang}.

In 1995, Jacobson examined the correlation between quantum corrections to the spacetime structure and black hole thermodynamics \cite{Ted}. The Anti-de-Sitter (AdS) spacetime is a background of constant negative curvature where the negative cosmological constant can be assumed as positive thermodynamic pressure \cite{Teitelboim}. AdS spacetime is believed to be a proper background since black holes can present well-defined thermal features, such as stability \cite{Hubeny}. In an interesting paper, Kastor et al. showed that the mass of a black hole in an AdS spacetime could be treated as an enthalpy function of its background \cite{Kastor}. For these and other reasons, there are many studies in the literature that frequently use AdS spacetime as the background \cite{Cham1, Cham2, XNWu, Kubi, Lir1, Lir2, DLi, BQWang, XYGUO, ZMXu, SWWei, ZGao, Okcu1, Okcu2, Eom}.

Unfortunately, black holes are objects that cannot be directly observed. Modern telescopes can only reveal their existence by detecting their effects on other adjacent matters. Astronomers obtained the first black hole image in 2019 with the Event Horizon Telescope from the center of the Messier 87 galaxy \cite{M871}. After this publication, which attracted great interest, in 2022, they published the image of Sagittarius A*, a supermassive black hole located at the heart of the Milky Way galaxy \cite{SagA1}. The common feature of both images is shadows surrounded by bright rings \cite{Chen2023}. The intense gravity of black holes deflects the light emitted by an illuminating source of light in their background \cite{He}. Therefore, spacetime properties, the size of the black hole, and the observers' relative position affect the shape of observed shadows. For example, a stationary observer would expect a perfect circle shadow for a Schwarzschild black hole \cite{Synge}, but in a rotating black hole, this expectation changes to a non-circular structure \cite{Bardeen}. The detection of black hole shadows has led many researchers to work on the theoretical modeling of black hole shadows in recent years \cite{Singh1, Kogan1, Kogan2, Konoplya, WeiMann, Babar, Kumar, Kumar1, Singh, LiGuo, Zhang21, Thomas1, Das, Volker, GuoWD, SS1, Heydar1, PG1, ZG1, Sunny1, Sunny2, Sunny3, Sunny4, Sunny5, Sunny6, Vir, Vir1}.

All these facts motivated us to investigate the shadow of a quantum-corrected SBH in AdS background surrounded by a quintessence matter field with its thermal quantities. To this end, in the following section, we first introduce the quantum-corrected SBH in AdS spacetime. Then, we extend the lapse function by considering the quintessence matter field and obtain the mass function. In section 3, we derive the thermal quantities. Next in section 4, we explore its shadow. After all, we briefly discuss the energy emission rate in section 5. Finally, we revisit our findings in the conclusion section to finalize the manuscript.

\section{Quantum-corrected Schwarzschild AdS black hole}

Wu and Liu explored the SBH in AdS spacetime due to spherically symmetric
quantum fluctuations in \cite{26}. There, they showed that the 4D theory of
gravity reduces to a two-dimensional dilaton gravity, and the modified
geometry of the quantum-corrected Schwarzschild-AdS black hole takes the
form of 
\begin{equation}
ds^{2}=-f\left( r\right) dt^{2}+\frac{1}{f\left( r\right) }%
dr^{2}+r^{2}\left( d\theta ^{2}+\sin \theta d\varphi ^{2}\right) ,  \label{L}
\end{equation}
where the lapse function stands for 
\begin{equation}
f\left( r\right) =\frac{1}{r}\sqrt{r^{2}-a^{2}}-\frac{2M}{r}+\frac{\left(
r^{2}-a^{2}\right) ^{3/2}}{\ell ^{2}r}.  \label{M}
\end{equation}
Here, $M$ corresponds to the mass of the quantum-corrected black hole, and $%
\ell $ represents the AdS radius. The deformation parameter, $a=4\ell _{p}$ (%
$\ell _{p}$ is the Planck length), explains the influence of spherical
symmetric quantum fluctuations. Since it sets a non-zero minimum value to $r$%
, the model becomes consistent with quantum gravity theories. If one takes
the deformation parameter as zero, then the modified metric reduces to the
SBH metric in the AdS background. Besides, in the limit of $\ell
^{2}\rightarrow \infty $, the quantum Schwarzschild AdS metric converts to
the usual Schwarzschild metric. We note that in the case of $r>>a$, the
quantum excitation effects are weak while the AdS modification influences
are strong. The opposite of this statement, i.e. AdS effects are weak and
quantum fluctuation effects are strong, is true when $r$ is approximately
larger than $a$.

When we expand the lapse function to the first order in $a^{2}$, we obtain 
\begin{equation}
f\left( r\right) \simeq \allowbreak 1-\frac{2M}{r}-\frac{a^{2}}{2r^{2}}+%
\frac{r^{2}}{\ell ^{2}}-\frac{3}{2}\frac{a^{2}}{\ell ^{2}}+\mathcal{O}\left(
a^{3}\right) .  \label{N}
\end{equation}%
We see that this form of the lapse function bears a resemblance to the lapse
function of the charged AdS spacetime.

\begin{figure}[tbh]
\begin{minipage}[t]{0.5\textwidth}
        \centering
        \includegraphics[width=\textwidth]{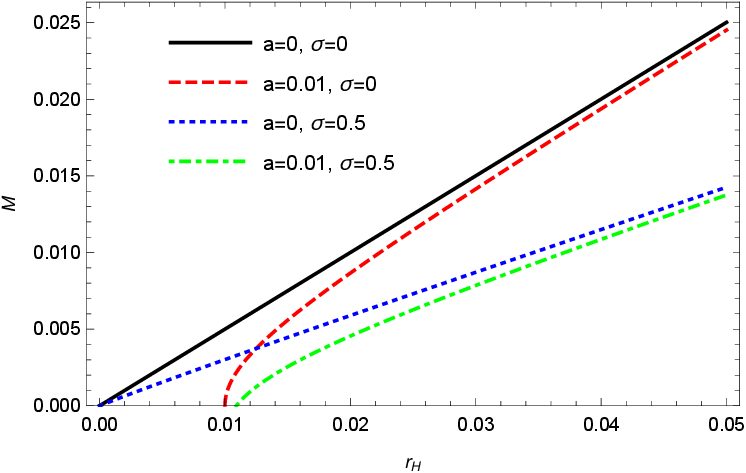}
       \subcaption{ $ \omega_q=-0.35$.}\label{fig:M1}
   \end{minipage}%
\begin{minipage}[t]{0.5\textwidth}
        \centering
       \includegraphics[width=0.96\textwidth]{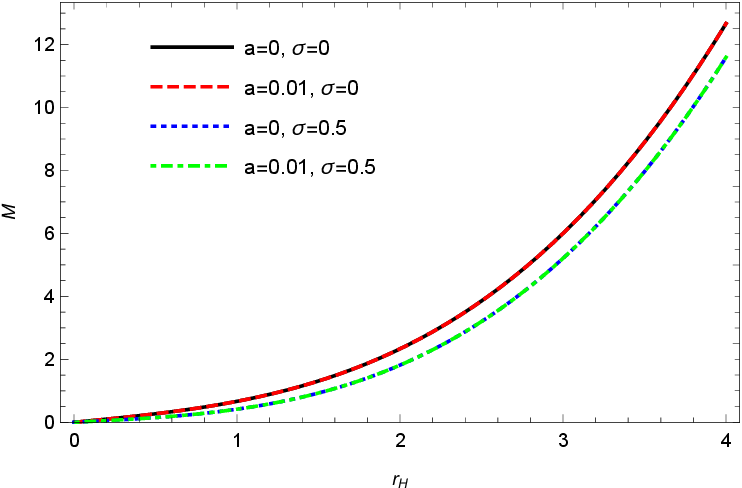}\\
        \subcaption{ $ \omega_q=-0.35$.}\label{fig:M1a}
    \end{minipage}\hfill 
\begin{minipage}[b]{0.5\textwidth}
        \centering
        \includegraphics[width=\textwidth]{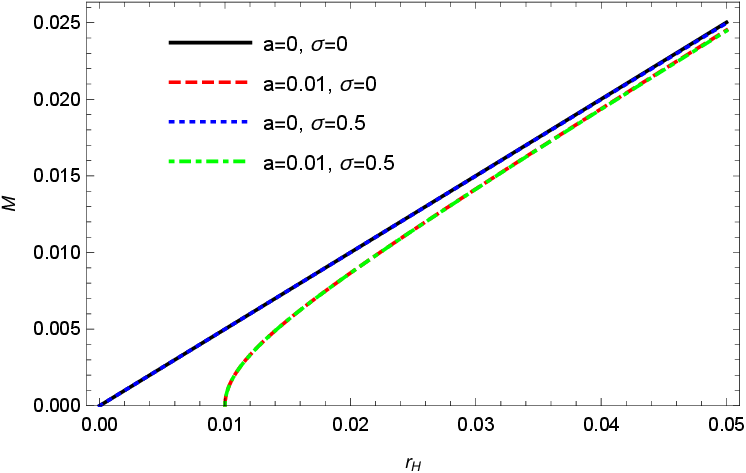}
       \subcaption{$ \omega_q=-0.95$.}\label{fig:M2}
   \end{minipage}%
\begin{minipage}[b]{0.5\textwidth}
        \centering
       \includegraphics[width=0.96\textwidth]{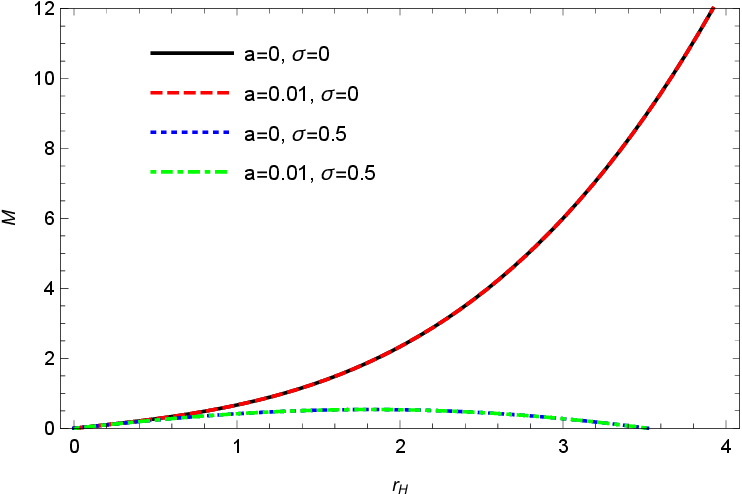}\\
        \subcaption{ $ \omega_q=-0.95$.}\label{fig:M2a}
    \end{minipage}\hfill
\caption{A qualitative comparison of the mass functions versus event horizon
for $P=1/8\protect\pi$.}
\label{Mfigsnew1}
\end{figure}

Now we consider the quantum-corrected SBH surrounded by a quintessence field
in the AdS background. To present the quintessence field we follow \cite%
{Shahjalal} and attach the term $\Big(-\frac{\sigma }{r^{3\omega_{q}+1}}\Big)
$ to the lapse function. 
\begin{equation}
f\left( r\right) =\frac{1}{r}\sqrt{r^{2}-a^{2}}-\frac{2M}{r}+\frac{\left(
r^{2}-a^{2}\right) ^{3/2}}{\ell ^{2}r}-\frac{\sigma }{r^{3\omega _{q}+1}}.
\label{met}
\end{equation}
Here, $\sigma $ denotes a normalization factor associated with the
quintessence matter field and $\omega _{q}$ corresponds to the state
parameter which satisfies the constraint, $\omega _{q}\in
\left(-1,-1/3\right)$. The roots of the modified lapse function correspond
to the spherical event horizons, $r_{H}$, 
\begin{equation}
\left. \frac{1}{r}\sqrt{r^{2}-a^{2}}-\frac{2M}{r}+\frac{\left(
r^{2}-a^{2}\right) ^{3/2}}{\ell ^{2}r}-\frac{\sigma }{r^{3\omega _{q}+1}}%
\right\vert _{r=r_{H}}=0,
\end{equation}
however, it is not possible to express the horizons with an analytical
expression due to the presence of undetermined quintessence matter
parameters. Nevertheless, we can express the mass function out of the lapse
function of the quantum-corrected Schwarzschild AdS black hole embedded in
the quintessence matter field as follows: 
\begin{equation}
M=\frac{1}{2}\sqrt{r_{H}^{2}-a^{2}}+\frac{4\pi P}{3}\left(
r_{H}^{2}-a^{2}\right) ^{3/2}-\frac{\sigma }{2r_{H}^{3\omega _{q}}},
\label{MASS}
\end{equation}
where the thermodynamic pressure is 
\begin{equation}
P=\frac{3}{8\pi \ell ^{2}}.
\end{equation}
For a detailed investigation, we depict the mass function versus the horizon
radius with different parameter values in Figs. \ref{Mfigsnew1} and \ref%
{Mfigsnew2}. 

\begin{figure}[tbh]
\begin{minipage}[t]{0.5\textwidth}
        \centering
        \includegraphics[width=\textwidth]{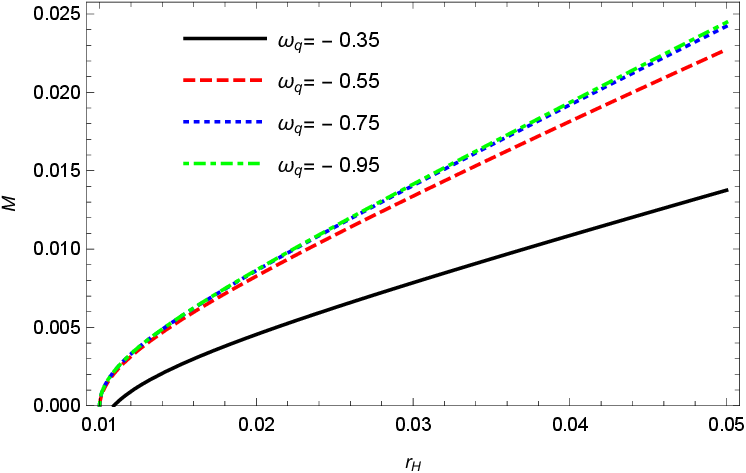}
       \subcaption{  $P=1/8\pi$.}\label{fig:M3}
   \end{minipage}%
\begin{minipage}[t]{0.50\textwidth}
        \centering
       \includegraphics[width=\textwidth]{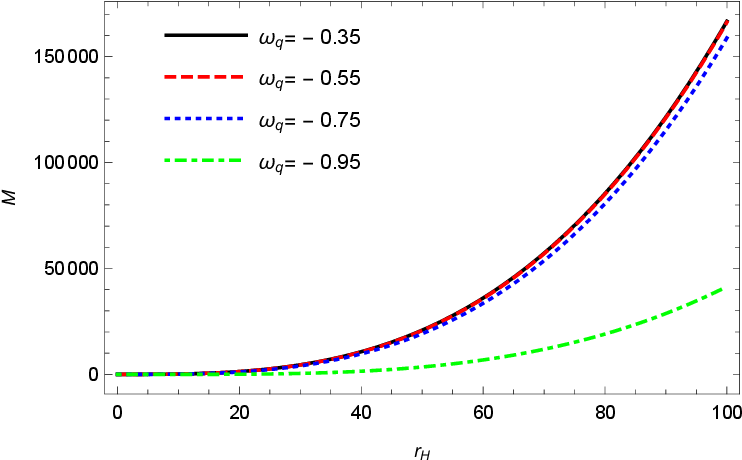}\\
        \subcaption{  $P=1/8\pi$.}\label{fig:M3a}
    \end{minipage}\hfill 
\begin{minipage}[b]{0.5\textwidth}
        \centering
        \includegraphics[width=\textwidth]{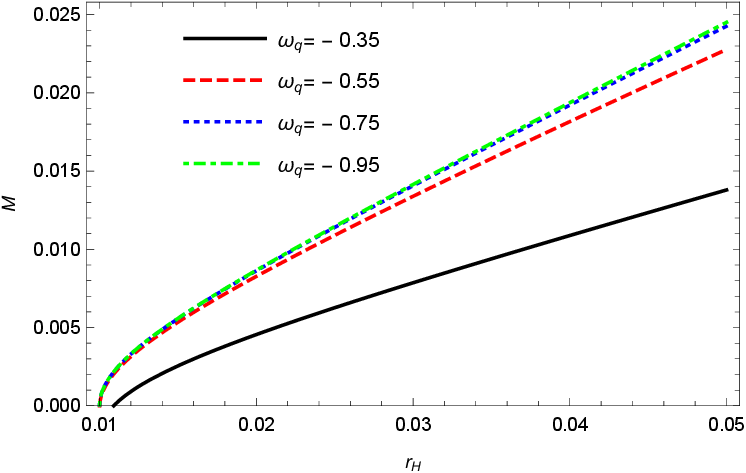}
       \subcaption{ $P=3/8\pi$.}\label{fig:M4}
   \end{minipage}%
\begin{minipage}[b]{0.5\textwidth}
        \centering
       \includegraphics[width=0.96\textwidth]{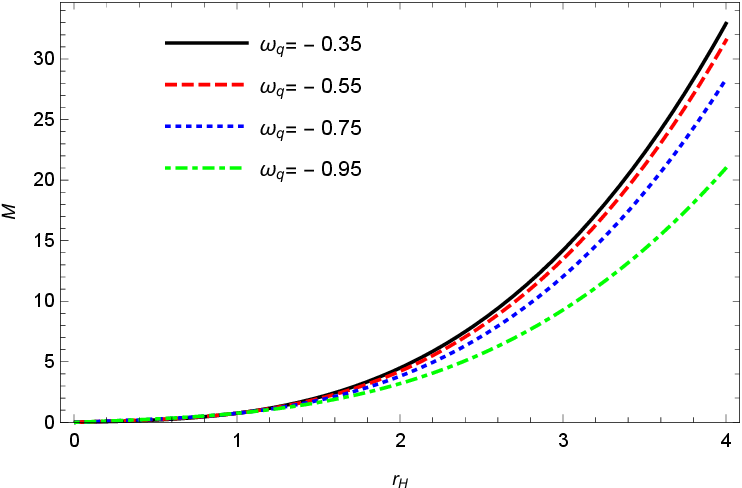}\\
        \subcaption{ $P=3/8\pi$.}\label{fig:4a}
    \end{minipage}\hfill
\caption{The effect of the quintessence state parameter on the mass function
for $a=0.01$ and $\protect\sigma=0.5$.}
\label{Mfigsnew2}
\end{figure}

In Figs. \ref{fig:M1} and \ref{fig:M2}, we see the minimal horizon value
which is aroused by the quantum correction. We observe that quantum
corrections are strong at relatively small event horizon values, while we
notice that quintessence matter effects are strong at relatively great
horizon values. We also note that the quintessence state parameter alters
the characteristic behavior of the mass function. E.g. for $\omega_q=-0.35$
the mass function only monotonically increases, while for $\omega_q=-0.95$
it does not. In the interval of $(0.01, 1.81692)$, the mass function
increases, and then, in between $( 1.81692, 3.5217)$ it decreases to zero.
In this case, the mass function takes its maximal value, $0.537039$, at $%
1.81692$. To have a better understanding of the effect of the quintessence
matter we plot Fig. \ref{Mfigsnew2}.

\newpage 
We observe that the mass takes smaller values at relatively small radii with
greater quintessence state parameters. This fact changes oppositely in
relatively greater radii. We see that if the AdS radius is greater enough
than the quintessence state parameter leads to a critical change in the mass
function. For $\omega_q=-0.95$, at great horizon radius values, the mass
function also monotonically increases. In order to find the start of this
increase, we solve Eq. \ref{MASS}. We find that the mass function starts to
take positive values for $r_H > 13.3552$. Therefore, we conclude that in
this case, the black hole cannot exist in the interval of $(3.5217,13.3552)$.

\section{Thermal quantities and critical values}

\begin{figure}[tbh]
\begin{minipage}[t]{0.5\textwidth}
        \centering
        \includegraphics[width=\textwidth]{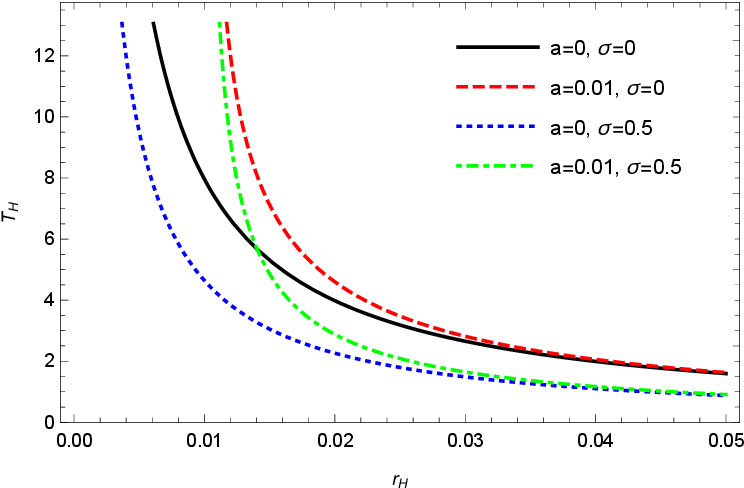}
       \subcaption{ $ \omega_q=-0.35$.}\label{fig:T1}
   \end{minipage}%
\begin{minipage}[t]{0.5\textwidth}
        \centering
       \includegraphics[width=\textwidth]{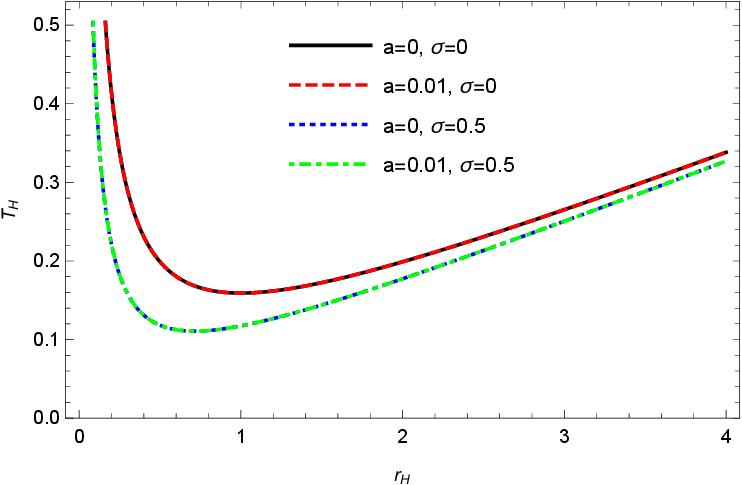}\\
        \subcaption{ $ \omega_q=-0.35$.}\label{fig:T1a}
    \end{minipage}\hfill 
\begin{minipage}[b]{0.5\textwidth}
        \centering
        \includegraphics[width=\textwidth]{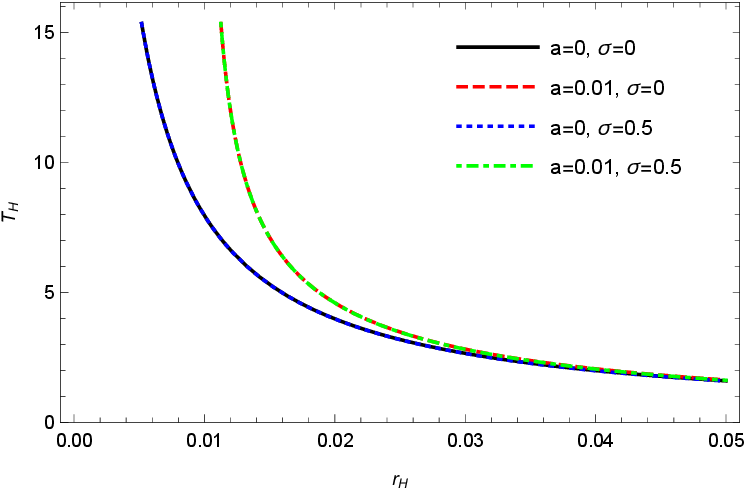}
       \subcaption{$ \omega_q=-0.95$.}\label{fig:T2}
   \end{minipage}%
\begin{minipage}[b]{0.5\textwidth}
        \centering
       \includegraphics[width=\textwidth]{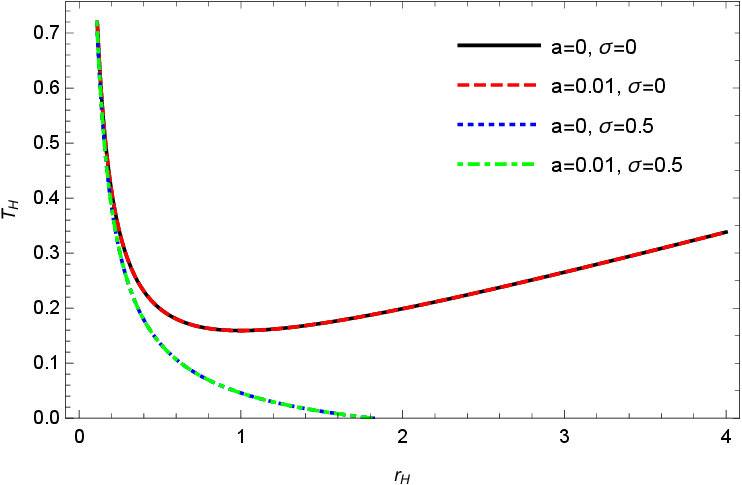}\\
        \subcaption{ $ \omega_q=-0.95$.}\label{fig:T2a}
    \end{minipage}\hfill
\caption{A qualitative comparison of the Hawking temperature versus event
horizon for $P=1/8\protect\pi$.}
\label{Tfigsnew1}
\end{figure}

\begin{figure}[tbh]
\begin{minipage}[t]{0.5\textwidth}
        \centering
        \includegraphics[width=\textwidth]{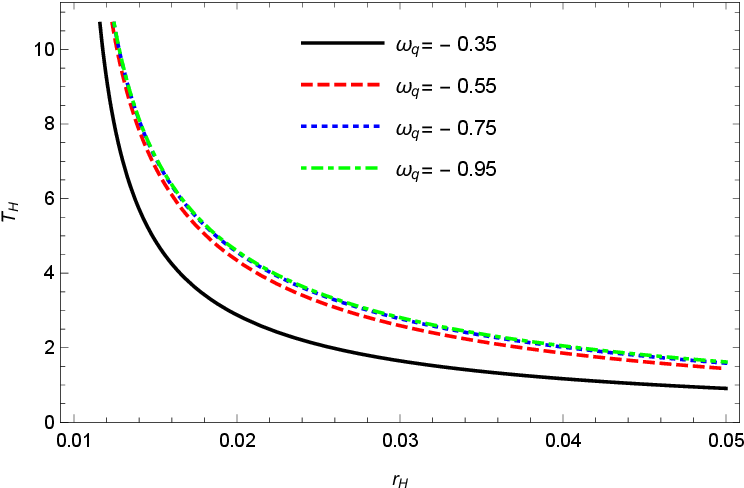}
       \subcaption{  $P=1/8\pi$.}\label{fig:T3}
   \end{minipage}%
\begin{minipage}[t]{0.50\textwidth}
        \centering
       \includegraphics[width=\textwidth]{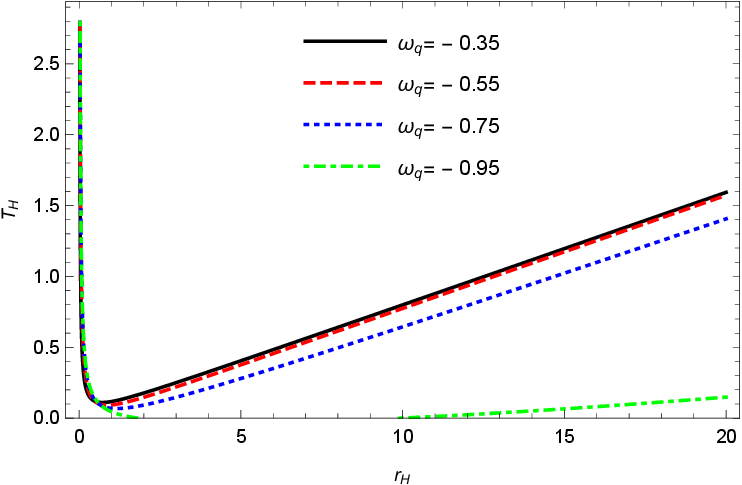}\\
        \subcaption{  $P=1/8\pi$.}\label{fig:T3a}
    \end{minipage}\hfill 
\begin{minipage}[b]{0.5\textwidth}
        \centering
        \includegraphics[width=\textwidth]{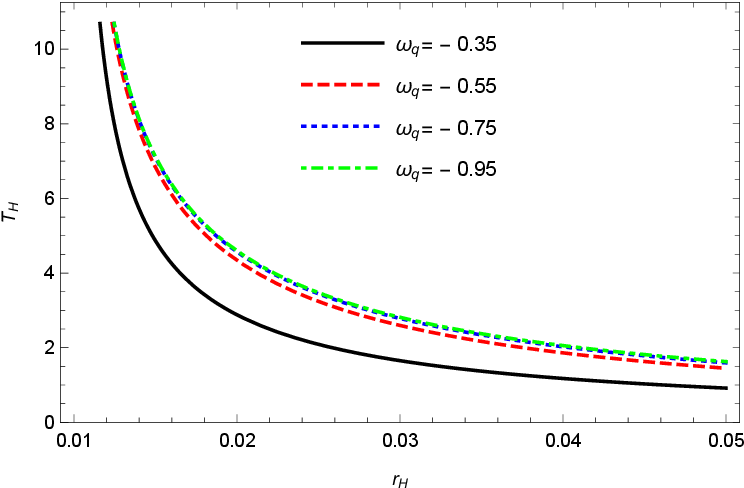}
       \subcaption{ $P=3/8\pi$.}\label{fig:T4}
   \end{minipage}%
\begin{minipage}[b]{0.5\textwidth}
        \centering
       \includegraphics[width=\textwidth]{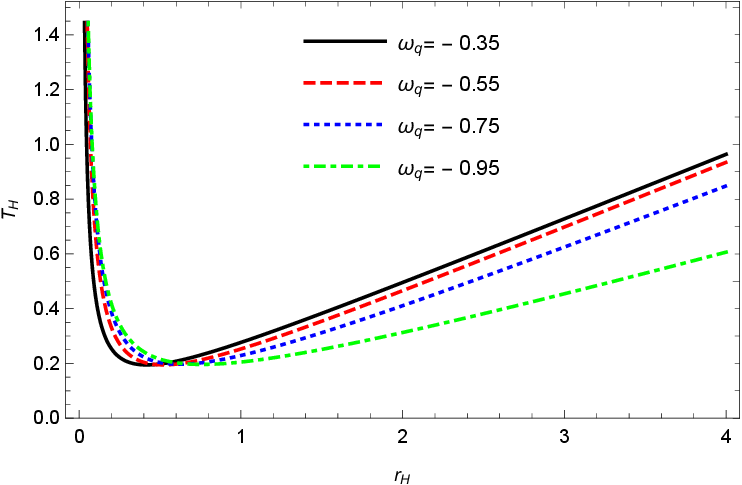}\\
        \subcaption{ $P=3/8\pi$.}\label{fig:T4a}
    \end{minipage}\hfill
\caption{The effect of the quintessence state parameter on the Hawking
temperature for $a=0.01$ and $\protect\sigma=0.5$.}
\label{Tfigsnew2}
\end{figure}

In this section, we first focus on the thermodynamics of the
quantum-corrected SBH surrounded by quintessence matter in the AdS
background. To this end, we initially examine the Hawking temperature with
the following the definition 
\begin{equation}
T_{H}=\frac{\kappa }{4\pi },  \label{t}
\end{equation}%
which utilizes the surface gravity, $\kappa$, through the existing formula 
\begin{equation}
\kappa =\left. \frac{d}{dr}f\left( r\right) \right\vert _{r=r_{H}}.
\end{equation}
Considering Eqs. (\ref{met}) and (\ref{MASS}), we obtain the modified
Hawking temperature as 
\begin{equation}
T_{H}=\frac{1}{4\pi \sqrt{r_{H}^{2}-a^{2}}}+2P\sqrt{r_{H}^{2}-a^{2}}+\frac{%
3\omega \sigma }{4\pi r_{H}^{3\omega _{q}+2}}.  \label{T}
\end{equation}
It is worth noting that for $a=0$, Eq. \eqref{T} reduces to the Hawking
temperature of the Schwarzschild AdS black hole surrounded by quintessence 
\begin{equation}
T_{H}=\frac{1}{4\pi r_{H}}+2Pr_{H}+\frac{3\omega \sigma }{4\pi
r_{H}^{3\omega _{q}+2}},
\end{equation}%
while by setting $\sigma =0$, it converts to the Hawking temperature of the
quantum-corrected Schwarzschild AdS black hole \cite{26} 
\begin{equation}
T_{H}=\frac{1}{4\pi \sqrt{r_{H}^{2}-a^{2}}}+2P\sqrt{r_{H}^{2}-a^{2}}.
\end{equation}%
Furthermore, in the case of $\sigma =a=0,$ Eq. \eqref{T} turns to the
ordinary Hawking temperature of the Schwarzschild AdS black hole \cite{Eom}, 
\begin{equation}
T_{H}=\frac{1}{4\pi r_{H}}+2Pr_{H}.
\end{equation}
To display these results, we plot the Hawking temperature versus event
horizon in Fig. \ref{Tfigsnew1}. \newpage 
Remembering that the Hawking temperature can only have real and positive
values, we impose the constraint, $r_{H}^{2}-a^{2}>0$. This restriction
establishes a lower bound value on the event horizon radius 
\begin{equation}
r_{H}>4\ell _{p},
\end{equation}
so that, a correlation between the event horizon radius and the Planck
length scale is established. Essentially, we see this condition in Figs. \ref%
{Mfigsnew1} and \ref{Tfigsnew1}. To observe the effects of the quintessence
matter parameter, we depict the Hawking temperature for two different
pressure values in Fig. \ref{Tfigsnew2}.

We observe that at relatively small event horizon values the Hawking
temperature takes greater values for smaller quintessence state parameters.
This fact changes after a certain horizon value and at relatively greater
horizon radii the Hawking temperature takes smaller values for smaller
quintessence state parameters. Besides, in Fig. \ref{fig:T2a} for a
particular value of the quintessence matter parameter, $\omega_q=-0.95$, the
Hawking temperature tends to zero at $r_H=1.81692$. In addition to this
fact, in Fig. \ref{fig:T3a}, we observe that the Hawking temperature starts
to take positive values after a certain horizon radius, $r_H=9.91074$. So,
we conclude that at the event horizon interval, $(1.81692,9.91074)$, the
black cannot exist. This result points out the presence of a set of critical
quintessence parameters.

Combining the two analyses for $\omega_q=-0.95$, $\sigma=0.5$, $P=1/8\pi$, $%
a=0.01$, we see that, unlike the other cases, the event horizon is bounded
in two intervals, $r_H \in (0.01, 1.81692)$ and $r_H> 13.3552$.

\newpage Subsequently, we employ the first law of black hole thermodynamics
to investigate the entropy 
\begin{equation}
S=\int \frac{dM}{T}.
\end{equation}
We obtain 
\begin{equation}
S=\pi r_{H}^{2},  \label{S}
\end{equation}
We see that the quantum correction and quintessence matter does not alter
the usual expression of the entropy function. However, it is clear that they
affect the event horizon radii and therefore the entropy in closed form.

\begin{figure}[tbh]
\begin{minipage}[t]{0.5\textwidth}
        \centering
        \includegraphics[width=\textwidth]{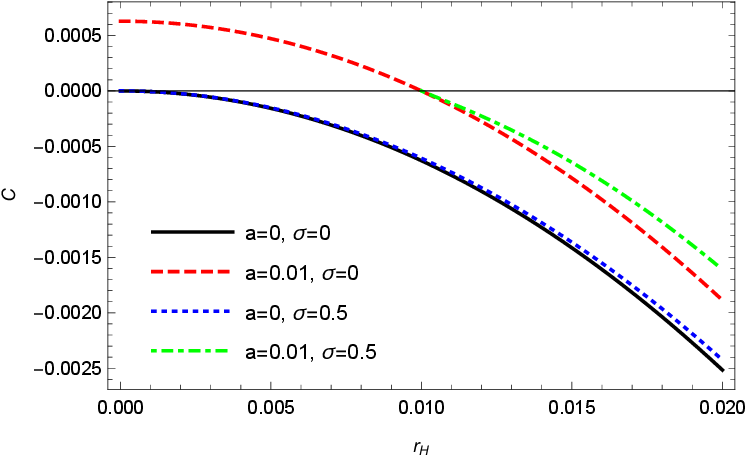}
       \subcaption{ $ \omega_q=-0.35$.}\label{fig:C1}
   \end{minipage}%
\begin{minipage}[t]{0.5\textwidth}
        \centering
       \includegraphics[width=\textwidth]{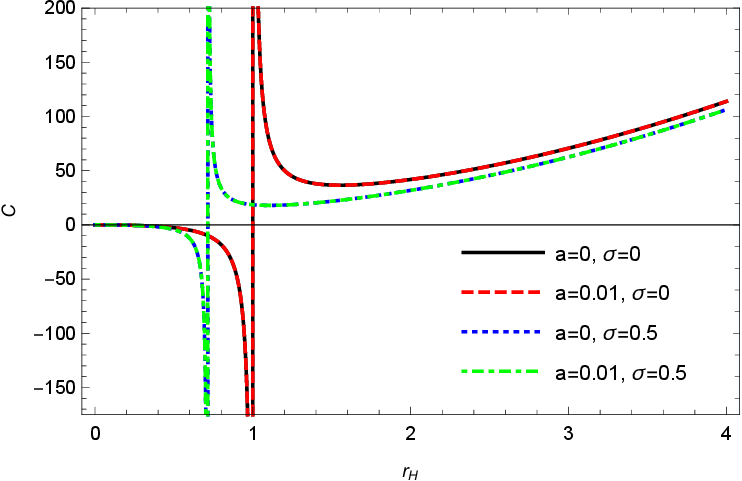}\\
        \subcaption{ $ \omega_q=-0.35$.}\label{fig:C1a}
    \end{minipage}\hfill 
\begin{minipage}[b]{0.5\textwidth}
        \centering
        \includegraphics[width=\textwidth]{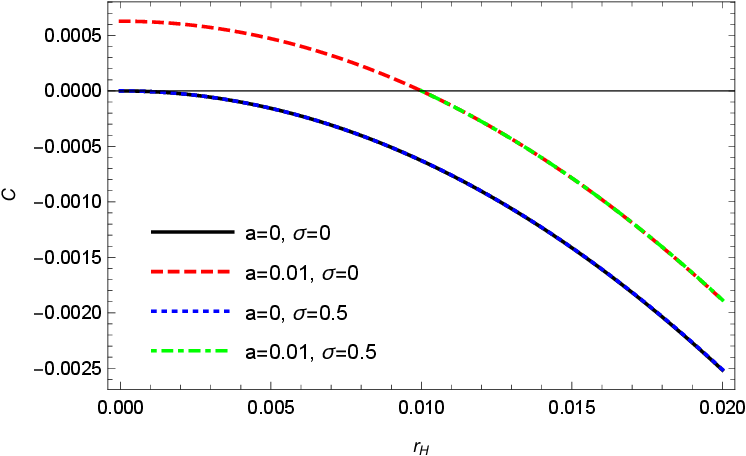}
       \subcaption{$ \omega_q=-0.95$.}\label{fig:C2}
   \end{minipage}%
\begin{minipage}[b]{0.5\textwidth}
        \centering
       \includegraphics[width=\textwidth]{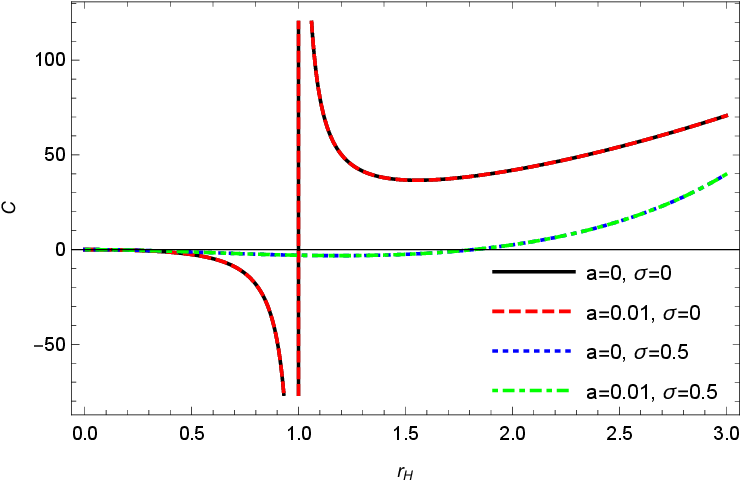}\\
        \subcaption{ $ \omega_q=-0.95$.}\label{fig:C2a}
    \end{minipage}\hfill
\caption{A qualitative comparison of the heat capacity function versus event
horizon for $P=1/8\protect\pi$.}
\label{Cfigsnew1}
\end{figure}

Next, we study the heat capacity function to investigate the thermal
stability of the black hole. In the black hole thermodynamics context,
negative values of the heat capacity indicate instability, while positive
values point out stability. We consider the formula 
\begin{equation}
C=\frac{dM}{dT},
\end{equation}%
to derive the heat capacity. By using Eqs. (\ref{MASS}) and (\ref{T}), we
achieve 
\begin{equation}
C=-2\pi \Bigg(\frac{r_{H}}{\sqrt{r_{H}^{2}-a^{2}}}+8\pi Pr_{H}\sqrt{%
r_{H}^{2}-a^{2}}+\frac{3\sigma \omega _{q}}{r_{H}^{3\omega _{q}+1}}\Bigg) %
\Bigg[\frac{r_{H}}{\left( r_{H}^{2}-a^{2}\right) ^{3/2}}-\frac{8\pi Pr_{H}}{%
\sqrt{r_{H}^{2}-a^{2}}}+\frac{3\sigma (3\omega _{q}+2)\omega _{q}}{%
r_{H}^{3\omega _{q}+3}}\Bigg]^{-1}.  \label{qSBHads specheat}
\end{equation}%
Before demonstrating our findings, we briefly would like to discuss several
particular cases. At first, we set $\sigma =0$. In this case, we obtain the
heat capacity of the quantum-corrected Schwarzschild AdS black hole in the
form of 
\begin{equation}
C=-2\pi \left( r_{H}^{2}-a^{2}\right) \frac{1+8\pi
P\left(r_{H}^{2}-a^{2}\right) }{1-8\pi P\left( r_{H}^{2}-a^{2}\right) },
\end{equation}%
which states that the black hole is not thermodynamically unstable for $%
r_{H}\in \left] a,\sqrt{a^{2}+\frac{1}{8\pi P}}\right[ $, while it is stable
for $r_{H}>\sqrt{a^{2}+\frac{1}{8\pi P}}$. For $\ell \rightarrow \infty$, we
find the quantum-corrected SBH specific heat function, $C=-2\pi \left(
r_{H}^{2}-a^{2}\right)$, which obviously reduces to the usual formula in the
absence of quantum-correction term. On the other hand, by setting $a=0$ in
Eq. \eqref{qSBHads specheat}, we get the heat capacity of the Schwarzschild
AdS black hole surrounded by quintessence matter 
\begin{equation}
C=-2\pi r_{H}^{2}\bigg(1+8\pi Pr_{H}^{2}+\frac{3\sigma \omega _{q}}{%
r_{H}^{3\omega _{q}+1}}\bigg)\bigg[1-8\pi Pr_{H}^{2}+\frac{3\sigma \omega
_{q}(3\omega _{q}+2)}{r_{H}^{3\omega _{q}+1}}\bigg]^{-1}.  \label{heat}
\end{equation}%
Here, we cannot reveal the thermal stability without determining particular
quintessence parameter values. Next, we can focus on the graphical analysis
of Eq. \eqref{qSBHads specheat}. At first, we present a qualitative
comparison in Fig. \ref{Cfigsnew1}.

\newpage We observe that for the case $\omega_q=-0.95$, the SBH is unstable
for $r_H \in (0.01, 1.81692)$. A remnant mass, $M_{rem}=0.537039$, occurs at 
$1.81692$. Then, in order to see the effect of the quintessence parameter,
we depict the heat capacity function versus the horizon radius with
different parameter values of pressure and $\omega _{q}$ in Fig. \ref%
{Cfigsnew2}.

\newpage 
\begin{figure}[tbh]
\begin{minipage}[t]{0.5\textwidth}
        \centering
        \includegraphics[width=\textwidth]{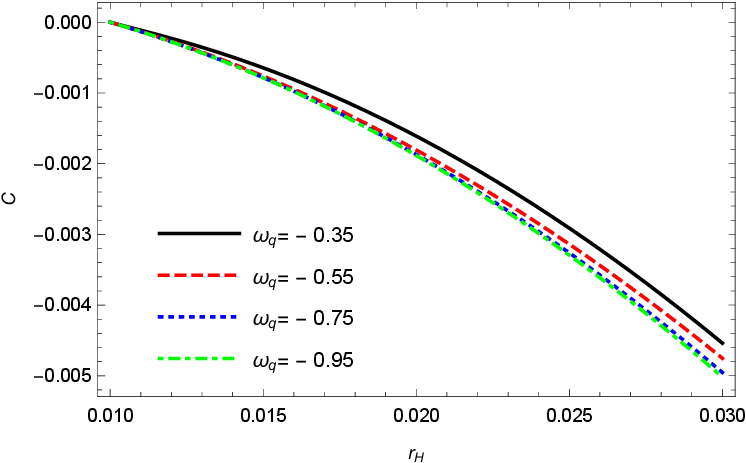}
       \subcaption{  $P=1/8\pi$.}\label{fig:C3}
   \end{minipage}%
\begin{minipage}[t]{0.50\textwidth}
        \centering
       \includegraphics[width=\textwidth]{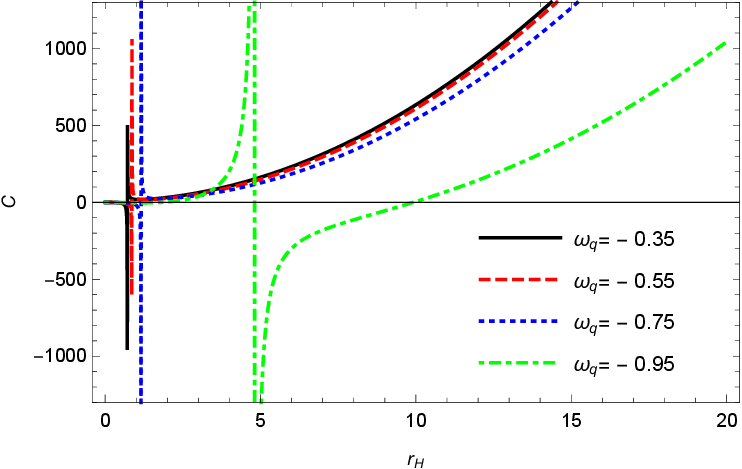}\\
        \subcaption{  $P=1/8\pi$.}\label{fig:C3a}
    \end{minipage}\hfill 
\begin{minipage}[b]{0.5\textwidth}
        \centering
        \includegraphics[width=\textwidth]{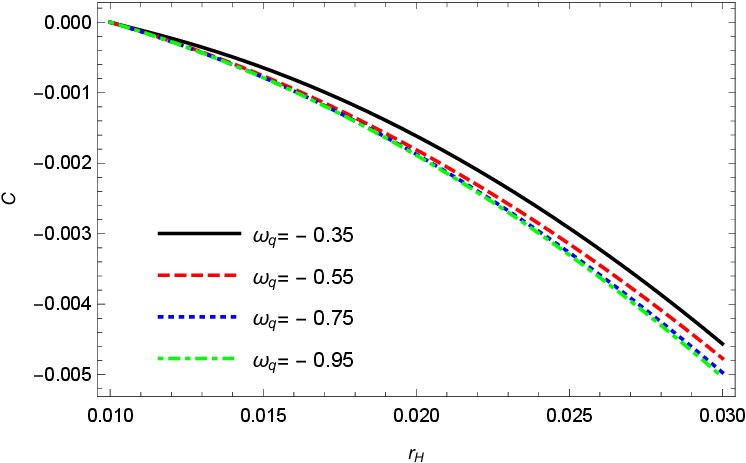}
       \subcaption{ $P=3/8\pi$.}\label{fig:C4}
   \end{minipage}%
\begin{minipage}[b]{0.5\textwidth}
        \centering
       \includegraphics[width=\textwidth]{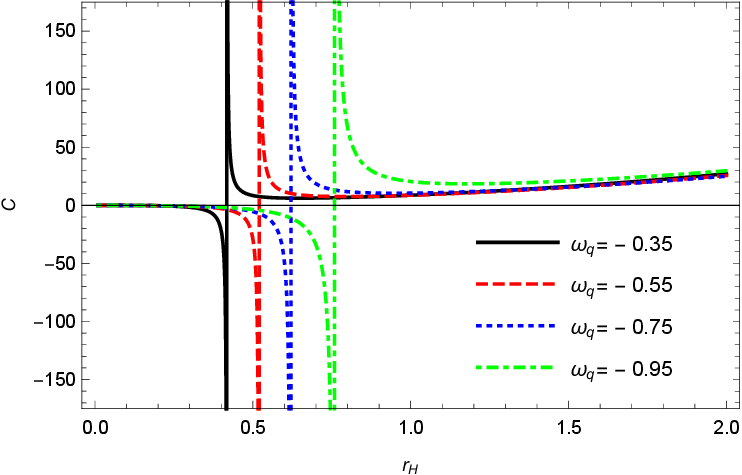}\\
        \subcaption{ $P=3/8\pi$.}\label{fig:C4a}
    \end{minipage}\hfill
\caption{The effect of the quintessence state parameter on the heat capacity
function for $a=0.01$ and $\protect\sigma=0.5$.}
\label{Cfigsnew2}
\end{figure}

We observe that in the case of $P=\frac{1}{8\pi}$ and $\omega _{q}=-0.95$,
the quantum-corrected SHB surrounded by the quintessence matter field in the
AdS background becomes stable only for $r_H> 13.3552$.

\begin{figure}[tbh]
\begin{minipage}[t]{0.35\textwidth}
        \centering
        \includegraphics[width=\textwidth]{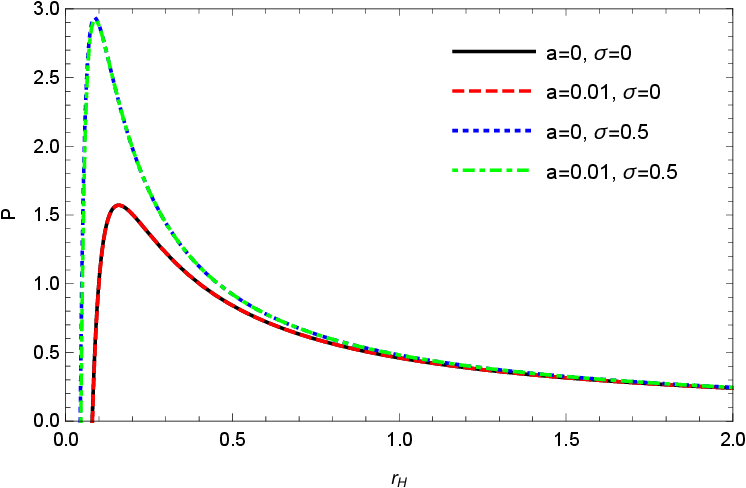}
       \subcaption{$T_H=1$, and  $ \omega_q=-0.35$.}\label{fig:Pa}
   \end{minipage}%
\begin{minipage}[t]{0.35\textwidth}
        \centering
        \includegraphics[width=\textwidth]{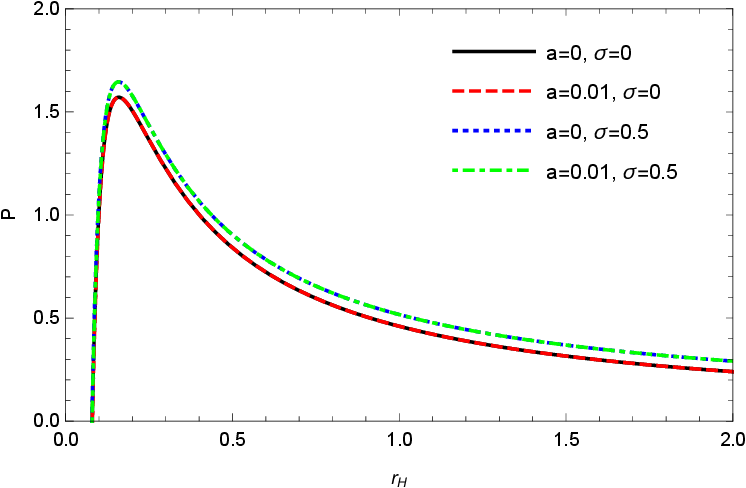}
       \subcaption{$T_H=1$, and  $ \omega_q=-0.95$.}\label{fig:Pb}
   \end{minipage}%
\begin{minipage}[t]{0.35\textwidth}
        \centering
        \includegraphics[width=\textwidth]{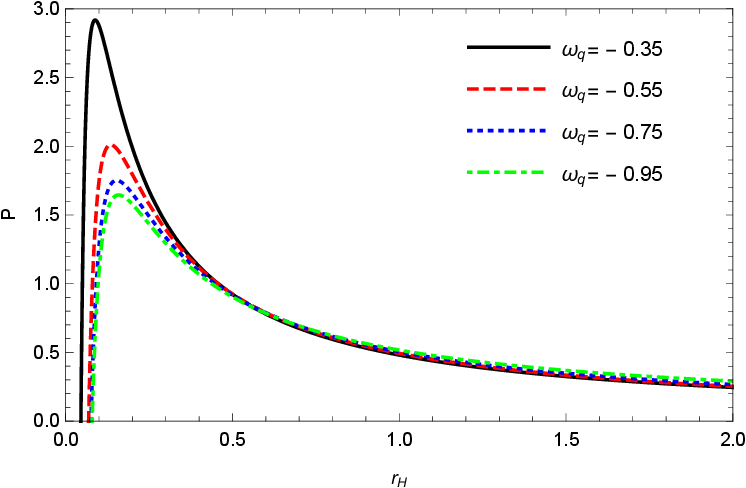}\\
        \subcaption{$T_H=1$, $a = 0.01 $,  and $\sigma = 0.5 $.}\label{fig:Pc}
    \end{minipage}\hfill
\caption{A qualitative comparison of black hole isotherms.}
\label{Iso}
\end{figure}

Now, let us investigate the critical thermodynamic variables and determine
the critical exponents. Using the derived Hawking temperature we write the
pressure as the function of temperature and horizon 
\begin{equation}
P=\frac{T}{2\sqrt{r_{H}^{2}-a^{2}}}-\frac{1}{8\pi \left(
r_{H}^{2}-a^{2}\right) }-\frac{3 \omega \sigma}{8\pi r_{H}^{3\omega +2} 
\sqrt{r_{H}^{2}-a^{2}}}.  \label{Pr}
\end{equation}
This function is called the geometric equation of state. In Fig. \ref{Iso},
we depict several isotherm functions to illustrate a comparison. In Fig. \ref%
{fig:Pa}, we observe the effects of quantum corrections at relatively small
horizons. For $\omega_q=-0.95$, we do not see this effect sharply. In Fig. %
\ref{fig:Pc}, we rank the isotherms of the quantum-corrected SBH according
to the $\omega_q$. We observe that the peak remains relatively smaller for
smaller values of the quintessence state parameter.

\newpage\ 

\section{Black Hole Shadow}

\begin{figure}[tbh]
\begin{minipage}[t]{0.5\textwidth}
        \centering
        \includegraphics[width=\textwidth]{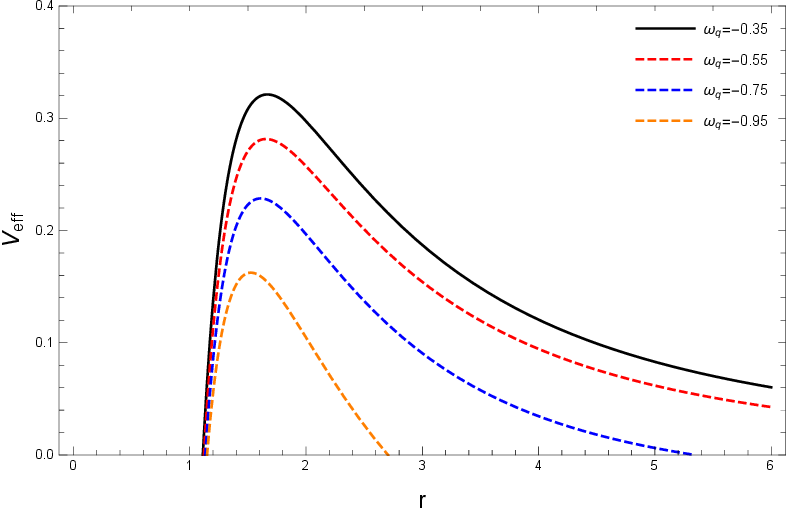}
       \subcaption{ $P=3/8 \pi$ and $L=1$.}\label{fig:Va}
   \end{minipage}%
\begin{minipage}[t]{0.50\textwidth}
        \centering
       \includegraphics[width=\textwidth]{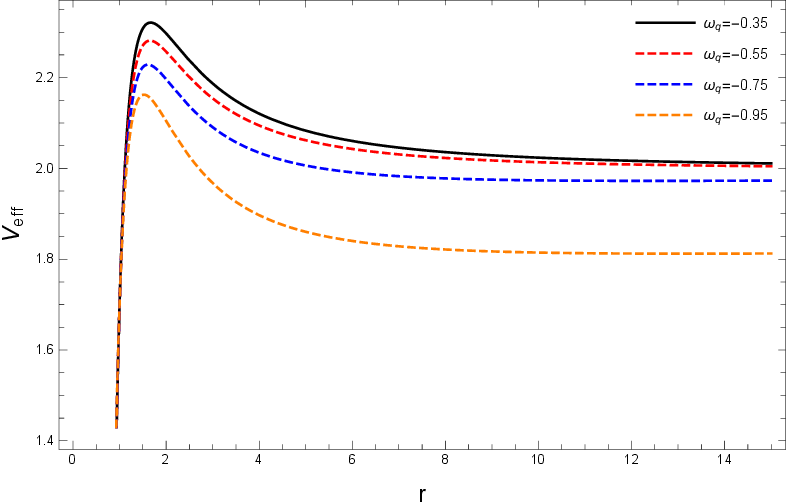}\\
        \subcaption{  $P=1/8 \pi$ and $L=1$.}\label{fig:Vb}
    \end{minipage}\hfill 
\begin{minipage}[b]{0.5\textwidth}
        \centering
        \includegraphics[width=\textwidth]{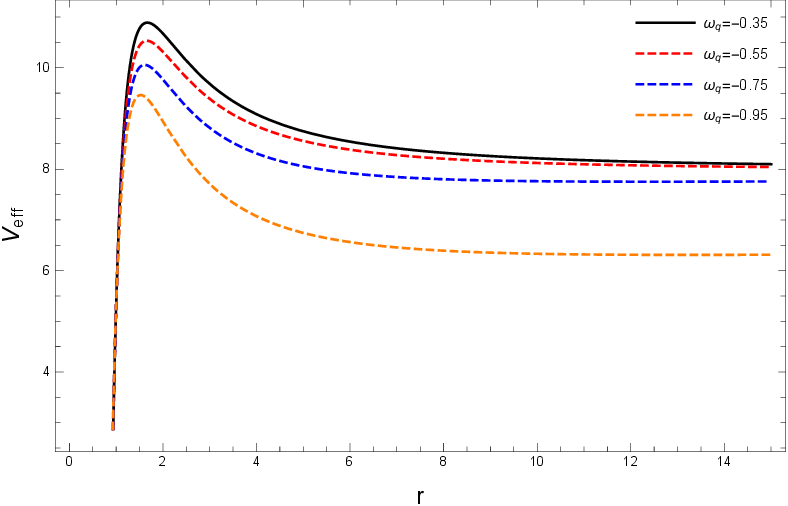}
       \subcaption{  $P=1/8 \pi$ and $L=5$.}\label{fig:Vc}
   \end{minipage}%
\begin{minipage}[b]{0.50\textwidth}
        \centering
       \includegraphics[width=\textwidth]{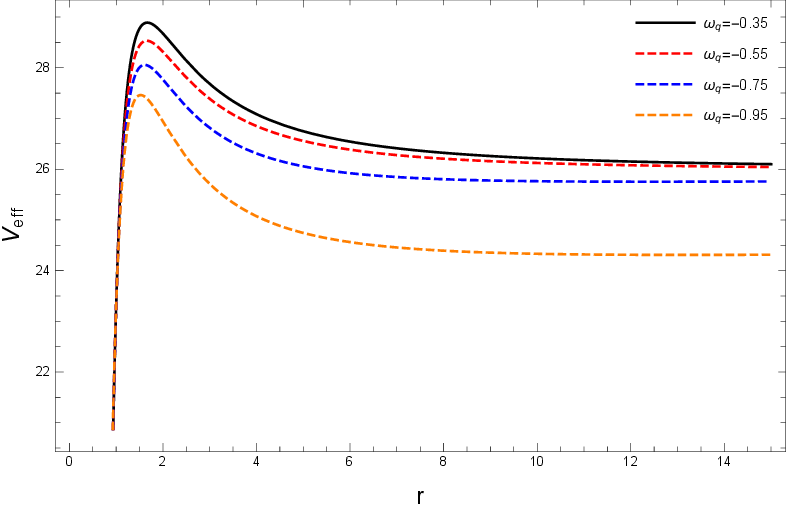}\\
        \subcaption{  $P=3/8 \pi$ and $L=5$}\label{fig:Vd}

    \end{minipage}\hfill
\caption{The influence of the quintessence state parameter on the effective
potential for $M=0.5$, $E=1$.}
\label{Vfig}
\end{figure}

When there is a black hole between the light source and the observer, the
black hole's gravitational field deflects the light. In such a situation,
some photons may fall into the black hole, creating a region devoid of light
known as the black hole shadow. The apparent shape of a black hole can be
predicted by the boundary of this shadow. The form of the shadow depends on
certain black hole parameters. Recent studies also reveal that the
quintessence matter field may alter the shape of the shadows. In this
section, we intend to model the shadow of the quantum-corrected
Schwarzschild AdS black hole that is embedded in the quintessence matter
field. We start with writing the Lagrangian for light rays 
\begin{equation}
\mathcal{L}=\frac{1}{2}g_{\mu \nu }\dot{x}^{\mu }\dot{x}^{\nu }.
\end{equation}
Here, $g_{\mu \nu }$ denotes the metric tensor and the dot over the
variables stands for the derivative with respect to the affine parameter, "$%
\tau $". In a spherically symmetric spacetime, we get the Lagrangian with
the following form 
\begin{equation}
\mathcal{L}=\frac{1}{2}\left[ -f\left( r\right) \dot{t}^{2}+\frac{1}{f\left(
r\right) }\dot{r}^{2}+r^{2}\left( \dot{\theta}^{2}+\sin ^{2}\theta \dot{%
\varphi}^{2}\right) \right] .
\end{equation}%
Then, we employ the lapse function of the considered quantum-corrected
Schwarzschild AdS black hole and obtain the canonically conjugate momentum
components, $P_{\mu }=\frac{\partial \mathcal{L}}{\partial \dot{x}^{\mu }}$,
as follows: 
\begin{equation}
P_{t}=\left( \frac{1}{r}\sqrt{r^{2}-a^{2}}-\frac{2M}{r}+\frac{\left(
r^{2}-a^{2}\right) ^{3/2}}{\ell ^{2}r}-\frac{\sigma }{r^{3\omega _{q}+1}}%
\right) \dot{t}=E,
\end{equation}
\begin{equation}
P_{r}=\left( \frac{1}{r}\sqrt{r^{2}-a^{2}}-\frac{2M}{r}+\frac{\left(
r^{2}-a^{2}\right) ^{3/2}}{\ell ^{2}r}-\frac{\sigma }{r^{3\omega _{q}+1}}%
\right) ^{-1}\dot{r},
\end{equation}%
\begin{equation}
P_{\theta }=r^{2}\dot{\theta},
\end{equation}%
\begin{equation}
P_{\varphi }=r^{2}\sin ^{2}\theta \dot{\varphi}=L.
\end{equation}
Here, $E$ and $L$ correspond to the energy and angular momentum,
respectively. In order to examine the geodesics motion in this spacetime, we
adopt the Carter separability prescription of the Hamilton-Jacobi equation
which has the following general form%
\begin{equation}
\frac{\partial }{\partial \tau }\mathcal{S}=-\frac{1}{2}g_{\mu \nu }\frac{%
\partial \mathcal{S}}{\partial x^{\mu }}\frac{\partial \mathcal{S}}{\partial
x^{\nu }}.  \label{S1}
\end{equation}%
By considering the following separable solution of the Jacobi action 
\begin{equation}
\mathcal{S}=-Et+\mathcal{S}_{r}\left( r\right) +\mathcal{S}_{\theta }\left(
\theta \right) +L\varphi,  \label{S2}
\end{equation}%
we get {\small 
\begin{equation}
\left( \sqrt{r^{2}-a^{2}}-2M+\frac{\left( r^{2}-a^{2}\right) ^{3/2}}{\ell
^{2}}-\frac{\sigma }{r^{3\omega _{q}}}\right) ^{2}\left( \frac{\partial S_{r}%
}{\partial r}\right) ^{2}=E^{2}-\frac{\mathcal{K}+L^{2}}{r^{2}}\left( \sqrt{%
r^{2}-a^{2}}-2M+\frac{\left( r^{2}-a^{2}\right) ^{3/2}}{\ell ^{2}}-\frac{%
\sigma }{r^{3\omega _{q}}}\right) ,\,\,\,\,  \label{255}
\end{equation}%
}%
\begin{equation}
\left( \frac{\partial S_{\theta }}{\partial \theta }\right) ^{2}=\mathcal{K}%
-L^{2}\cot ^{2}\theta .  \label{28}
\end{equation}
Then, we recast Eq. (\ref{255}) and Eq. (\ref{28}) as 
\begin{equation}
r^{2}\left( \frac{\partial r}{\partial \tau }\right) =\pm \sqrt{\mathcal{R}},
\label{530}
\end{equation}%
\begin{equation}
r^{2}\left( \frac{\partial \theta }{\partial \tau }\right) =\pm \sqrt{\Theta 
},
\end{equation}
with 
\begin{equation}
\mathcal{R}=E^{2}\left[ r^{4}-r^{2}\left( \frac{1}{r}\sqrt{r^{2}-a^{2}}-%
\frac{2M}{r}+\frac{\left( r^{2}-a^{2}\right) ^{3/2}}{\ell ^{2}r}-\frac{%
\sigma }{r^{3\omega _{q}+1}}\right) \left( \eta +\zeta ^{2}\right) \right] ,
\end{equation}
and 
\begin{equation}
\Theta =E^{2}\left[ \eta -\zeta ^{2}\cot ^{2}\theta \right] .
\end{equation}
Here, $\eta =\frac{\mathcal{K}}{E^{2}}$ and $\zeta =\frac{L}{E}$ stands for
the impact parameters. It is widely recognized that the unstable null
circular orbits can be used for obtaining the shape of the shadow cast. To
this end, we rewrite Eq.~(\ref{530}) in the following form%
\begin{equation}
\left( \frac{\partial r}{\partial \tau }\right) ^{2}+V_{eff}\left( r\right)
=0,
\end{equation}
where the effective radial potential reads 
\begin{equation}
V_{eff}\left( r\right) =\left[ \frac{\mathcal{K}+L^{2}}{r^{2}}\left( \frac{1%
}{r}\sqrt{r^{2}-a^{2}}-\frac{2M}{r}+\frac{\left( r^{2}-a^{2}\right) ^{3/2}}{%
\ell ^{2}r}-\frac{\sigma }{r^{3\omega _{q}+1}}\right) -E^{2}\right] ,
\end{equation}
or%
\begin{equation}
V_{eff}\left( r\right) =E^{2}\left[ \frac{\eta +\zeta ^{2}}{r^{2}}\left( 
\frac{1}{r}\sqrt{r^{2}-a^{2}}-\frac{2M}{r}+\frac{\left( r^{2}-a^{2}\right)
^{3/2}}{\ell ^{2}r}-\frac{\sigma }{r^{3\omega _{q}+1}}\right) -1\right] .
\label{Veff}
\end{equation}
In Fig.(\ref{Vfig}) we demonstrate the behavior of the effective potential
function versus $r$ for different values of $\sigma $, $\omega _{q}$ and $a$%
. We observe that the effective potential takes greater values with larger
values of the quintessence state parameter. We see a similar effect for the
peak values of the effective potential, too.

The effective potential's maximum value corresponds to the circular orbits,
and the unstable photons should satisfy the following condition 
\begin{equation}
\left. V_{eff}\left( r\right) \right\vert _{r=r_{p}}=\left. \frac{d}{dr}%
V_{eff}\left( r\right) \right\vert _{r=r_{p}}=0,  \label{VL}
\end{equation}
or alternatively 
\begin{equation}
\left. \mathcal{R}\left( r\right) \right\vert _{r=r_{p}}=\left. \frac{d}{dr}%
\mathcal{R}\left( r\right) \right\vert _{r=r_{p}}=0.  \label{RP}
\end{equation}%
Here, $r_{p}$ is the radius of the photon sphere. From the Eqs. (\ref{VL})
and (\ref{RP}), we find that $r_{p}$ is the solution of the following
equation 
\begin{equation}
r_{p}f^{\prime }\left( r_{p}\right) -2f\left( r_{p}\right) =0.  \label{psr}
\end{equation}
By using Eq. (\ref{RP}), we find the following condition 
\begin{equation}
\eta +\zeta ^{2}=\frac{4r_{p}^{2}}{\frac{2r_{p}^{2}+a^{2}}{r_{p}\sqrt{%
r_{p}^{2}-a^{2}}}-\frac{2M}{r_{p}}+\frac{8\pi P}{3r_{p}}\sqrt{r_{p}^{2}-a^{2}%
}\left( 4r_{p}^{2}-a^{2}\right) +\frac{\sigma \left( 3\omega _{q}-1\right) }{%
r_{p}^{3\omega _{q}+1}}},
\end{equation}%
which should be satisfied by the impact parameters. Unfortunately, it is
quite difficult to solve Eq. (\ref{psr}) analytically, however, numerical
methods can be employed. In Table (\ref{tab:week3}), we give a set of
solutions for the quantity $\eta +\zeta ^{2}$ for three different values of $%
\sigma$. It is worth emphasizing that the solutions also obey the condition, 
$\left. \frac{d^{2}}{dr^{2}}V_{eff}\left( r\right) \right\vert_{r=r_{p}}<0$. 
\begin{table}[htbp]
\centering
\begin{tabular}{l|ll|ll|ll}
\hline\hline
& $\sigma =0.1$ &  & $\sigma =0.2$ &  & $\sigma =0.3$ &  \\ \hline
$\omega _{q}$ & $r_{p}$ & $\eta +\zeta ^{2}$ & $r_{p}$ & $\eta +\zeta ^{2}$
& $r_{p}$ & $\eta +\zeta ^{2}$ \\ \hline
-0.35 & 1.66705 & 0.903335 & 1.87834 & 0.931108 & 2.15555 & 0.953973 \\ 
-0.45 & 1.66443 & 0.908354 & 1.89002 & 0.941781 & 2.23208 & 0.969926 \\ 
-0.55 & 1.65536 & 0.914263 & 1.88435 & 0.95502 & 2.30172 & 0.990722 \\ 
-0.65 & 1.63768 & 0.921196 & 1.84812 & 0.971472 & 2.29869 & 1.01864 \\ 
-0.75 & 1.60976 & 0.929249 & 1.77205 & 0.991598 & 2.09335 & 1.05622 \\ 
-0.85 & 1.57156 & 0.938445 & 1.66559 & 1.01522 & 1.80437 & 1.10272 \\ 
-0.95 & 1.52527 & 0.948685 & 1.55288 & 1.04129 & 1.58372 & 1.15357 \\ \hline
\end{tabular}%
\caption{The values of the photon radius, $r_{p}$, and $\protect\eta +%
\protect\zeta ^{2}$ for different values of $\protect\sigma $ with $a=0.01$, 
$M=0.5$ and $P=3/8\protect\pi $. }
\label{tab:week3}
\end{table}

Our findings reveal that for a fixed value of $\omega _{q}$, the quantity $%
\eta +\zeta ^{2}$ increases by greater values of the normalization factor, $%
\sigma $. Moreover, the quantity $\eta +\zeta ^{2}$ increases generally with
the decreasing values of the quintessence state parameter.

\begin{figure}[tbh]
\begin{minipage}[t]{0.5\textwidth}
        \centering
        \includegraphics[width=\textwidth]{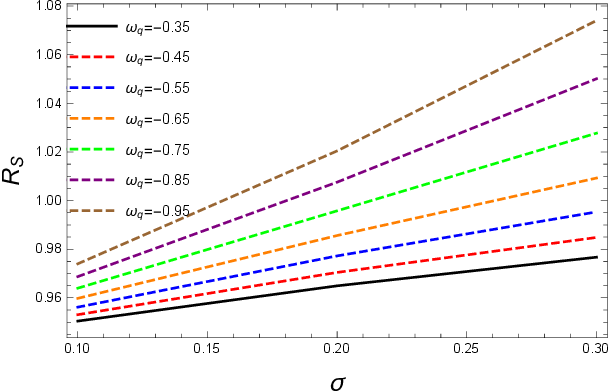}
      
   \end{minipage}%
\begin{minipage}[t]{0.50\textwidth}
        \centering
       \includegraphics[width=\textwidth]{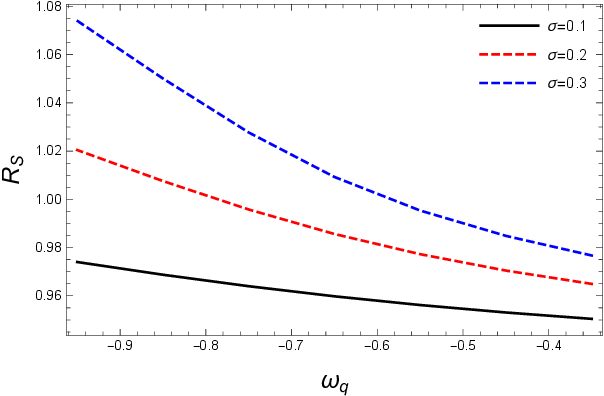}\\
       
    \end{minipage}\hfill
\caption{The radius of black hole shadow versus quintessence matter field
parameters for $P=3/8\protect\pi$, $M=0.5$ and $a=0.01$.}
\label{Shfig}
\end{figure}

In \cite{Vazque}, authors reported that using the celestial coordinates $X$
and $Y$ is more suitable to describe the real shadow of the black hole seen
on the observer's frame. According to \cite{Singh}, the celestial
coordinates can be defined with: 
\begin{equation}
X=\lim_{r_{o}\rightarrow \infty }\left( -r_{o}\sin \theta _{o}\frac{d\varphi 
}{dr}\right) ,  \label{XX}
\end{equation}%
\begin{equation}
Y=\lim_{r_{o}\rightarrow \infty }\left( r_{o}^{2}\frac{d\theta }{dr}\right) ,
\label{YY}
\end{equation}
where $r_{o}$ denotes the distance between the black hole and far observer,
while $\theta _{o}$ stands for the angular position of the observer with
respect to the black hole's plane. We consider the observer on the
equatorial hyperplane ($\theta=\pi /2$), so that we simplify Eqs.(\ref{XX})
and (\ref{YY}) to the following form of 
\begin{equation}
X^{2}+Y^{2}=\eta +\zeta ^{2}=R_{S}^{2}.
\end{equation}
Here, $R_{S}^{2}$ is the radius of the black hole shadow. In Fig. \ref{Shfig}%
, we present the plots of the shadow radius versus the normalization and
quintessence state parameters. 

\begin{figure}[htb]
\begin{minipage}[t]{0.5\textwidth}
        \centering
        \includegraphics[width=\textwidth]{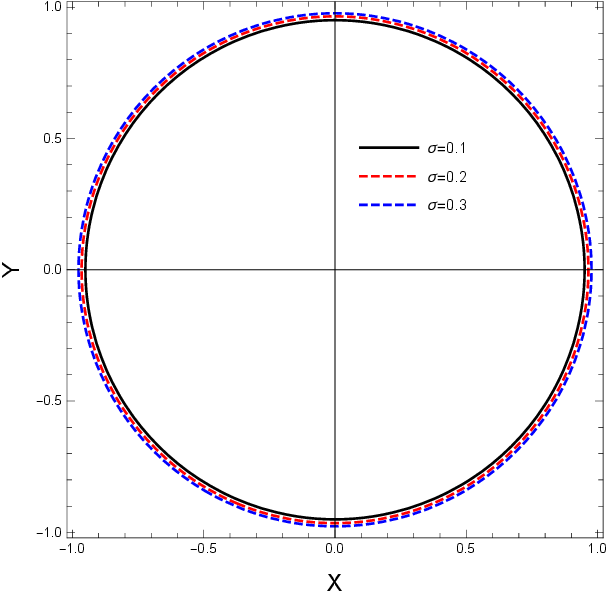}
       \subcaption{ $ \omega_q=-0.35$.}\label{fig:Spa}
   \end{minipage}%
\begin{minipage}[t]{0.50\textwidth}
        \centering
       \includegraphics[width=\textwidth]{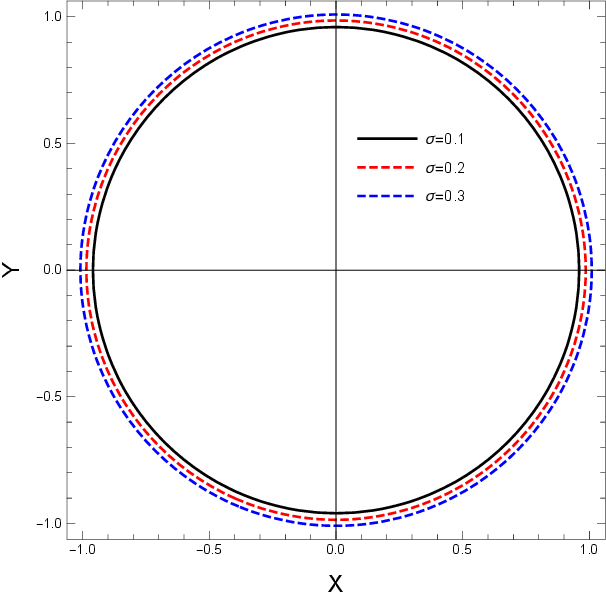}\\
        \subcaption{  $ \omega_q=-0.65$.}\label{fig:Spb}
    \end{minipage}\hfill 
\begin{minipage}[b]{0.5\textwidth}
        \centering
        \includegraphics[width=\textwidth]{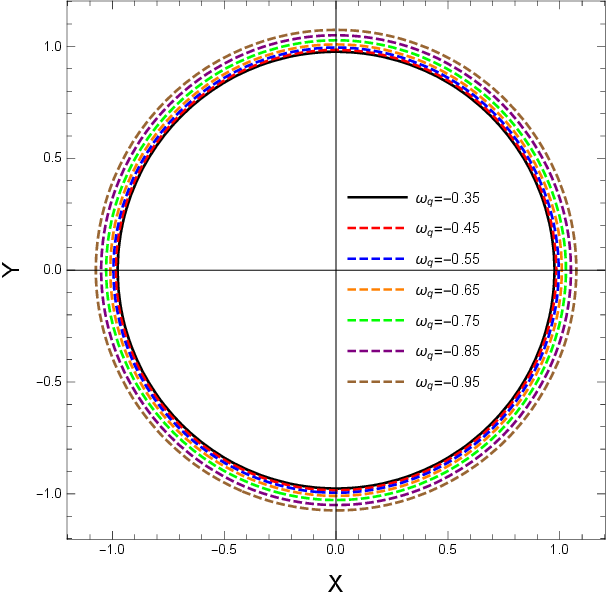}
       \subcaption{ $\sigma=0.3$.}\label{fig:Spc}
   \end{minipage}%
\begin{minipage}[b]{0.50\textwidth}
        \centering
       \includegraphics[width=\textwidth]{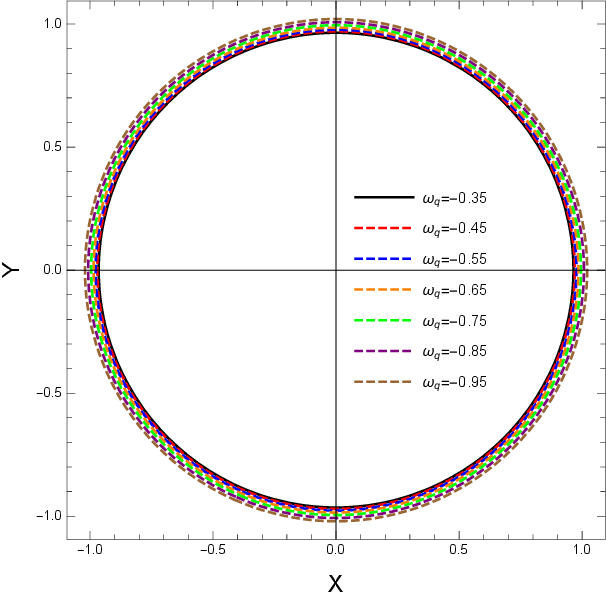}
       \subcaption{ $\sigma=0.2$.}\label{fig:Spd}
    \end{minipage}\hfill
\caption{Shadow of quantum-corrected Schwarzschild AdS black hole for
different values of quintessence matter parameters for $P=3/8\protect\pi$, $%
M=0.5$ and $a=0.01$. }
\label{Spfig}
\end{figure}

\newpage We observe that an increase in the $\sigma$ parameter leads to an
increase in the size of the black hole shadow. On the other hand, for a
fixed $\sigma$ value, the size of the shadow decreases by increasing values
of the quintessence parameter $\omega_{q}$. Finally, we display the
stereographic projection of the black hole's shadow in In Fig. \ref{Spfig}.

We see that the quintessence matter field varies the size of the black hole
shadow in the observer's frame.

\section{Energy emission rate}

Before we conclude the manuscript, we want to investigate the energy
emission rate. According to \cite{Ennadifi}, it can be calculated via the
formula 
\begin{equation}
\frac{d^{2}E}{d\omega dt}=\frac{2\pi ^{2}\sigma _{\lim }}{e^{\frac{\omega }{%
T_H}}-1}\omega ^{3}.  \label{em}
\end{equation}
Here, $T_H$ is the Hawking temperature, which is derived in Eq. (\ref{T}), $%
\omega$ is the emission frequency, and $\sigma _{\lim }$ is the absorption
cross-section, which is roughly equal to the geometrical cross-section of
the photon sphere. 
\begin{equation}
\sigma _{\lim }\sim \pi R_{S}^{2}.  \label{sig}
\end{equation}%
We substitute Eq. \eqref{sig} in Eq. \eqref{em}, and then, in Fig.~\ref%
{Efigsnew} we plot the variation of the energy emission rate versus
frequency for various quintessence state and normalization parameters
values. 
\begin{figure}[htb]
\begin{minipage}[t]{0.5\textwidth}
        \centering
        \includegraphics[width=\textwidth]{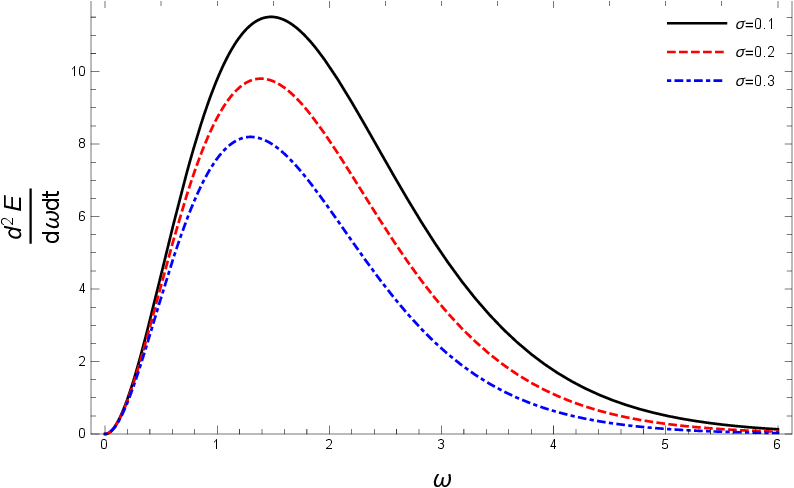}
       \subcaption{ $ \omega_q=-0.35$}\label{fig:EMa}
   \end{minipage}%
\begin{minipage}[t]{0.50\textwidth}
        \centering
       \includegraphics[width=\textwidth]{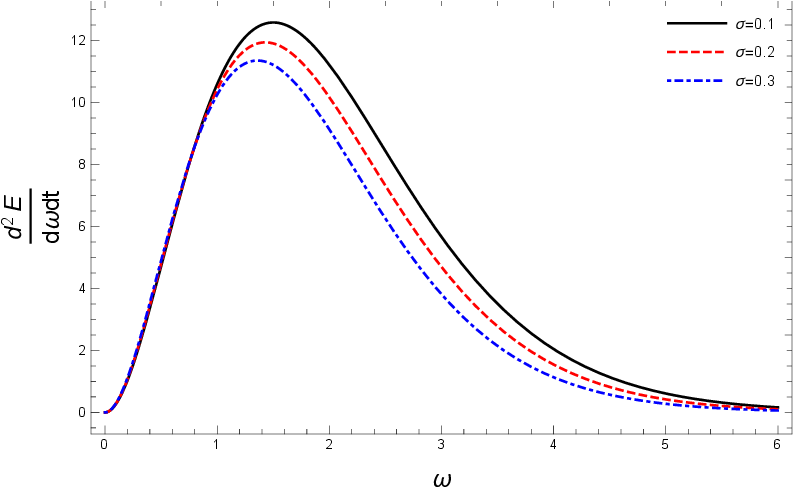}\\
        \subcaption{  $ \omega_q=-0.95$}\label{fig:EMb}
    \end{minipage}\hfill
\caption{Variation of energy emission rate versus frequency for $a=0.01$, $%
M=0.5$ and $P=3/8\protect\pi$.}
\label{Efigsnew}
\end{figure}

We observe that the peak value of that Gaussian-type plot decreases when the
normalization parameter gets greater values. Moreover, for smaller values of
the quintessence parameter, the peak occurs at greater frequencies.

\section{Conclusion}

In this manuscript, we explored the shadows and thermal quantities of a
quantum-corrected Schwarzschild AdS black hole surrounded by quintessence
matter. For this purpose, we attached the quintessence matter field terms to
the lapse function of the quantum-corrected Schwarzschild AdS black hole. At
first, we obtained the mass function, and we discussed the effects of the
quintessence matter field in the presence and absence of quantum
corrections. We observed that quantum corrections are effective only in
relatively small event horizon radii. We found that the quintessence field
effects are more effective on relatively greater event horizon radii. We
also found that for a particular value of the quintessence state parameter,
the black hole cannot exist for all event horizon values. Then, we derived
the Hawking temperature. Our detailed analysis revealed similar effects of
quantum corrections and quintessence matter fields. Then, we studied the
entropy function. We showed that its functional form does not change. After
that, we derived the specific heat function to discuss the stability of the
black hole. We found that the black hole could be stable or unstable,
depending on the event horizon value. Next, we found the geometric equation
of state and investigated the isotherms graphically. 

Then, we investigated the black hole's shadows. We found that the
quintessence matter field parameters alter the black hole shadow as follows:
An increase in the normalization factor increases the radius size of the
shadow. Moreover, at a fixed normalization parameter case, the size of the
shadow decreases for greater values of the quintessence state parameter.

Finally, we studied the energy emission rate and showed that the peak value
of the Gaussian-type function decreases for greater normalization parameter
values. Moreover, we demonstrated that the peak occurs at greater
frequencies for smaller values of the quintessence parameter.

\bigskip

\section*{Acknowledgments}

%The authors are thankful to the anonymous reviewer for the constructive comments. 
This manuscript is supported by the Internal Project, [2023/2211], of
Excellent Research of the Faculty of Science of Hradec Kr\'alov\'e
University.

\section*{Data Availability Statements}

The authors declare that the data supporting the findings of this study are
available within the article.

\end{document}